\documentclass[twocolumn,aps,superscriptaddress,showpacs]{revtex4}
\usepackage{times}
\usepackage{xspace}
\usepackage{graphicx}
\usepackage{color}
\usepackage{amsmath, amsthm, amssymb}

\newcommand{\bbold}[1]{\mbox{\boldmath$#1$}}
\newcommand{\phvec}{\ensuremath{\mathbf{k \varepsilon}}\xspace}
\newcommand{\phvecp}{\ensuremath{\mathbf{k' \varepsilon'}}\xspace}

\begin{document}

\title{Theory for wavelength-resolved photon emission statistics in 
single-molecule spectroscopy}
\author{Golan Bel \footnote{Present Address: Center for Nonlinear Studies, Los Alamos National Laboratory, Los Alamos, NM 87545 USA}} 

\affiliation{Department of Chemistry and Biochemistry and Department of Physics,
University of California, Santa Barbara, California 93106, USA}
\author{Frank~L.~H.~Brown}
\affiliation{Department of Chemistry and Biochemistry and Department of Physics,
University of California, Santa Barbara, California 93106, USA}

\begin{abstract}
We derive the moment generating function for photon emissions
from a single molecule driven by laser excitation.  
The frequencies of the fluoresced photons are explicitly considered.  
Calculations are performed for the case of a two level dye molecule,
showing that measured photon statistics will display a strong and 
non-intuitive dependence on detector bandwidth. Moreover, it is
demonstrated that the anti-bunching phenomenon, associated with 
negative values of Mandel's Q-parameter, results from correlations 
between photons with well separated frequencies.

\end{abstract}


\maketitle

Single-molecule spectroscopy (SMS) \cite{moernerrev} 
provides a detailed glimpse into our natural world.
Typically,  SMS experiments rely upon
broadband detection of fluoresced photons to monitor
molecular dynamics, however in studies of resonance
energy transfer (RET) \cite{spret} and semiconductor
quantum dots \cite{dot} considerably more information
can be obtained by resolving photon emission into
two color channels.  Recently \cite{luong}, 
multi-channel detection schemes have been 
introduced to extend the capabilities of SMS still further.

The information obtained by SMS is useful only 
if it can be readily interpreted.
SMS has received considerable
theoretical attention (see reviews \cite{silbrev} and references within), but 
most of this work ignores any consideration of photon color.  
A notable exception is the
treatment of RET, which has been considered in detail \cite{szabo}.  
A related treatment of ``frequency resolved"  photon counting, 
including quantum evolution of the molecule, has also been proposed 
by us \cite{bel}.  However, these studies rely upon a direct correspondence
between individual spectral transitions and experimental detection channels.  This
picture may be adequate for well resolved transitions and certain experimental
conditions, but falls short of providing a complete theoretical description of 
emission spectroscopy at the single-molecule level.  

Within the field of quantum optics,
time correlations between spectrally resolved photons have been studied both
experimentally \cite{CT2} and theoretically \cite{Nienhuis} for the case of resonance
fluorescence from 2-level atoms.  These studies also rely upon a direct
correspondence between individual spectral transitions (in the dressed-atom picture
\cite{ct}) and the frequency of the emitted photons to enable elementary interpretation
of experiment and simplified theoretical analysis.  This letter introduces a 
general formalism to describe single molecule photon emission that does
not presume simplifying characteristics of the molecular system or detection
apparatus.  Our results may be directly applied to model systems
and lay the groundwork for development of controlled approximation schemes 
in the study of more complex condensed-phase systems.

In previous work, we \cite{zheng1} and others \cite{cook,szabo} have introduced the generating
function formalism for calculation of single-molecule photon counting 
statistics without spectral resolution.  Such broadband photon statistics may 
be calculated by monitoring the number of times that spontaneous emission
occurs as the molecule evolves.  Within the Markovian limit
for molecular dynamics, spontaneous emission is a simple rate process
and these emission events may be treated purely classically, even though the
underlying dynamics may involve facets of
quantum evolution.  Calculation of photon counting moments proceeds via
introduction of the generating function for spontaneous emission events
$G(s,t)\equiv \langle s^{n(t)} \rangle$ where $n(t)$ is the number of emissions 
in the interval $[0,t]$ and
the factorial moments of this quantity follow immediately by differentiating 
$G$ with regard to the auxiliary variable $s$ and evaluating at $s=1$.  
The equations of motion for $G(s,t)$ (and by extension the factorial moments) 
involve only minimal complications beyond the usual quantum master equation 
approach used to solve for density matrix dynamics \cite{zheng1}.

In contrast to the above, if the frequency of emitted photons is measured, it becomes
impossible to proceed via simple classical arguments.  Decay of an electronic
excitation into a particular field mode or narrow subset of modes can not be
monitored by simply counting instantaneous spontaneous emission ``events"; such a process
is fundamentally non-Markovian. However, the definition of the generating function
may be generalized to allow for calculation of factorial moments with frequency
resolution by explicitly introducing a quantum mechanical description of the radiation field.  
We take
\begin{align}
\label{eq:gendef}
G(\vec{s},t) &\equiv \left \langle \exp \left [ \sum_{\phvec}\ln (s_{\phvec})a^{\dagger}_{\phvec}(t)a_{\phvec}(t) \right ] \right \rangle \\
&= \left \langle \mathcal{N} \exp \left [ \sum_{\phvec}(s_{\phvec}-1)a^{\dagger}_{\phvec}(t)a_{\phvec}(t) \right ] \right \rangle \nonumber.
\end{align}
Here, the averaging operation has its usual meaning $\langle \ldots \rangle \equiv \mathrm{Trace}
\{ ... \rho(0) \}$ involving the initial density matrix and a full trace over both the fluorophore and radiation 
field degrees of freedom. 
Creation and annihilation operators for photons with wavevector $\mathbf{k}$ and polarization
$\bbold{\varepsilon}$ have been introduced to express $s^{n(t)}$ from the broad-band definition as $\exp \left [\ln(s) \sum_{\phvec} N_{\phvec}(t) \right ]$ with $N_{\phvec}(t)= a^{\dagger}_{\phvec}(t)a_{\phvec}(t)$ representing
the Heisenberg picture number operator for each mode.  The generalization from $s$ to $\vec{s}$ has been made to
facilitate extraction of spectral information.  The second equality, involving the normal ordering operator $\mathcal{N}$, follows from standard operator identities \cite{wilcox}.  Taylor expanding both
expressions around $s_{\phvec}=1$ reveals that the multivariate factorial moments of the
number operators $N_{\phvec}$ are obtainable by the  traditional differentiation rule at
$s_{\phvec}=1$ and that these moments are most conveniently expressed as a normally
ordered product of creation and annihilation operators for each mode appearing in a given
moment.  For example, we find
\begin{align}
\label{eq:derivs}
& \left .\frac{\partial^{n+m}G(\vec{s},t)}{\partial s_{\phvec}^{n}\partial s_{\phvecp}^{m}}\right |_{\vec{s}=1} = \langle N_{\phvec}^{(n)}(t)N_{\phvecp}^{(m)}(t) \rangle  \\
& =  \langle [(a^{\dagger}_{\phvec})^n (a^{\dagger}_{\phvecp})^m (a_{\phvecp})^m (a_{\phvec})^n](t) \rangle \nonumber
\end{align}
with the expected generalization applying to moments involving more than two modes.
The above introduces the notation:  $N^{(m)}\equiv N(N-1)\ldots (N-m+1)$.

To make further progress, we specify the form of the Hamiltonian governing the time evolution of the
operators discussed above \cite{ct}.
\begin{equation}
\label{eq:ham}
H(t) = H_s + H_R + H_I - \left ( D^+ (\bbold{\mu}_0 \cdot \bbold{E}_L) \frac{e^{-i \omega_L t}}{2} + h.c. \right)
\end{equation}
$H_s$ is the Hamiltonian for the system (atom or molecule) of interest, which
will always be modeled with two electronic states (ground and excited) coupled 
to nuclear degrees of freedom.
$H_R = \sum_{\phvec} \hbar \omega_{\mathbf{k}}a^{\dagger}_{\phvec}a_{\phvec}$
is the Hamiltonian for the quantum radiation field ($\omega_k  = c k$ with $c$ the speed of
light).  The last term in parentheses reflects a semi-classical
coupling between the applied laser field (assumed monochromatic with frequency $\omega_L$ and
amplitude $E_L$) and the system
within the dipole approximation ($\mathbf{D}\equiv \bbold{\mu}_0(D^+ + D^-)$ is the dipole moment operator for the system consisting of terms that raise ($+$) and lower ($-$) the electronic state
of the system) and
rotating wave approximation (RWA) \cite{ct}.
$H_I$ describes the interaction between the system and the modes of the quantized
electromagnetic field, also within the RWA and dipole approximation
\begin{equation}
H_I =  \gamma \sum_{\phvec}\left( -i (\bbold{\varepsilon}  \cdot \bbold{\mu}_0)D^+  a_{\phvec} + h.c. \right).
\end{equation}
In the above $\gamma =  \sqrt{\frac{\hbar \omega_{0}}{2\epsilon V}}$ where $\omega_{0}$ is the transition frequency between excited and ground
electronic states \cite{note}, $\epsilon$ is the permittivity and 
$V$ is the volume of the cubic box used to quantize the field ($V \rightarrow \infty$
below and does not appear in any final results).

The Heisenberg equations of motion for the creation and annihilation operators
evolving with dynamics dictated by eq. \ref{eq:ham} may be formally integrated to yield \cite{ct}
\begin{equation}
a_{\phvec}\left(t\right)= e^{-i\omega_k t} \left [a_{\phvec}\left(0\right)+ \frac{\gamma}{\hbar} \int_{0}^{t} (\bbold{\varepsilon}\cdot \bbold{\mu}_0)\widetilde{D}^{-}\left(\tau\right)
e^{i\omega_{kL} \tau}d\tau \right ].\label{eq:casol}
\end{equation}
and the conjugate expression for $a^{\dagger}_{\phvec}(t)$.  For later convenience, we have
introduced the slowly varying rotating frame operators $\widetilde{D}^{\pm}(t) \equiv D^{\pm}(t)e^{\mp i\omega_L t}$ and have set $\omega_{kL}=\omega_k - \omega_L$. From this, it is readily
seen that $a_{\phvec}(t)$ commutes with $\dot{a}_{\phvec}(t)$ and similarly for
$a_{\phvec}^\dagger(t)$ and $\dot{a}_{\phvec}^{\dagger}(t)$.  This fact, along with
the assumption that the initial time total (system and radiation field) density matrix 
is a direct product between the system and the vacuum state for the field ( i.e. $\rho(0)=\sigma_s(0)\bigotimes|0\rangle\langle0|$) allows us to reformulate eq.
\ref{eq:gendef} as  \cite{inprep}
\begin{widetext}
\begin{equation}
G\left(t,\overrightarrow{s}\right)=\left\langle \mathcal{T}_{\mathcal{N}}\exp\left({\displaystyle  \frac{\gamma^2}{\hbar^2} \sum_{\phvec}(s_{\phvec}-1){\displaystyle \int\limits _{0}^{t}}{\displaystyle \int\limits _{0}^{t}}(\bbold{\varepsilon} \cdot \bbold{\mu}_0)^2 \widetilde{D}^{+}\left(u\right)\widetilde{D}^{-}\left(v\right)e^{-i\omega_{kL}\left(u-v\right)}dudv}\right)\right\rangle. \label{eq:GF2}
\end{equation}
\end{widetext}
The operator $\mathcal{T_N}$ acts on all operators to the right of it
by first arranging all ``+'' operators to the left of all ``-'' operators and subsequently
placing all ``-'' operators in standard time order (latest times at the left) and all
``+'' operators in reversed time order (latest times at the right).  The advantage of
eq. \ref{eq:GF2} over either expression in eq. \ref{eq:gendef} is that the generating
function is now defined solely in terms of the evolution of the system, which allows
us to pursue actual calculations as detailed below.

Eq. \ref{eq:gendef} provides a theoretical route toward arbitrary photon counting moments.
For simplicity and to make connection with possible experiments, we specialize to the
case that photon detection is insensitive to propagation direction and polarization of the emitted photons
and also assume that the detectors have finite resolution, registering the arrival of all photons 
within a window of width $\Delta$ around a central frequency $\omega$.  We define
a number operator for photons within this window 
\begin{equation}
N_{(\omega,\Delta)} = \sum\limits_{\phvec:(\omega - \Delta/2) \leq \omega_k \leq (\omega + \Delta/2)} N_{\phvec}
\end{equation}
Combining the above definition with eqs. \ref{eq:derivs} and \ref{eq:GF2} and proceeding to the
continuum limit ($V \rightarrow \infty$) leads to the conclusion that
\small
\begin{align}
\label{eq:moms}
& \langle N^{(m)}_{(\omega,\Delta)} (t) \rangle =   \\
& \left \langle \mathcal{T_N} \left ( \frac{\Gamma_0}{2\pi}\int\limits_{\omega-\Delta/2}^{\omega+\Delta/2}
d \omega_1 \int \limits_{0}^{t} \int\limits_0^t \widetilde{D}^{+}\left(u\right)\widetilde{D}^{-}\left(v\right)e^{-i\omega_{1L}\left(u-v\right)}dudv \right)^m  \right \rangle \nonumber
\end{align}
\normalsize
where $\Gamma_{0}\equiv\frac{\omega_{0}^{3}\left|\mu_{0}\right|^{2}}{3\pi\epsilon\hbar c^{3}}$.
Eq. \ref{eq:moms} applied to the case $m=1$ counts, on average, the number of photons within
a given frequency window emitted in time $t$ by the externally excited molecule.  The time derivative of this quantity evaluated in the $t\rightarrow \infty$ limit reproduces the usual 
expression \cite{mollow,ct} for the spectrum of fluoresced radiation.  Also, in the limit that $\Delta \rightarrow
\infty$, eq. \ref{eq:moms} reduces to Mandel's expression \cite{optlett} for the factorial moments of photon emission
as detected in broadband measurements (i.e. no frequency resolution).

The $\mathcal{T_N}$
operator in eq. \ref{eq:moms} insures that all correlation functions appearing in 
$\langle N^{(m)}_{(\omega,\Delta)} (t)\rangle$ are of the form
\begin{equation}
\langle \widetilde{D}^+ (u_1)\ldots\widetilde{D}^+ (u_m)\widetilde{D}^-(v_m)
\ldots\widetilde{D}^-(v_1) \rangle
\end{equation}
with $u_m \geq u_{m-1}\geq \ldots u_1$ and $v_m \geq v_{m-1}\geq \ldots v_1$.  Correlation
functions with such time ordering may be calculated within in the Markov limit for system dynamics \cite{ct}
via an extension of the quantum regression theorem \cite{gardiner_noise}.

It follows that the explicit calculation of moments in eq. \ref{eq:moms} is straightforward in principle, involving only 
diagonalization of the rotating-frame evolution operator for system dynamics and elementary
integrals over time and frequency.  In practice, however, the procedure is complicated and will be specified 
in detail elsewhere \cite{inprep}.

For concreteness, we present predictions for the low temperature spectroscopy of 
a single 2-level dye molecule.  The 2-level molecule is specified by $H_s=\left(\hbar\omega_0/2\right)\left(|e\rangle \langle e |-|g\rangle \langle g|\right) $
and $D^+ =  |e\rangle \langle g|$  ($D^- = |g\rangle \langle e|$) with $e$ and $g$ designating the excited
and ground states.  Traditionally, the spontaneous emission rate from the excited state $\Gamma_0$ and the Rabi frequency $\Omega = \bbold{E}_L\cdot \bbold{\mu}_0 / \hbar$ \cite{ct} are specified in lieu of $\bbold{\mu}_0$ and $\bbold{E_L}$ and we follow this convention here.  We take $\Gamma_0/2\pi =  40 MHz$ in all that follows to model the organic dye terrylene in a hexadecane Shpol'skii matrix at
$1.7^{\circ}$ K,
a prototypical 2-level SMS system \cite{MPI}.  The following calculations assume various 
values of $\Delta$, ranging from 0.2 MHz to 200 MHz.  Resolution down to 2 MHz is possible 
using a Fabry-Perot interferometer \cite{grove}.  Theoretically, it should be possible 
to measure most of the reported quantities.
%
%

\begin{figure}[t]
\centering
\includegraphics[width=3.4in]{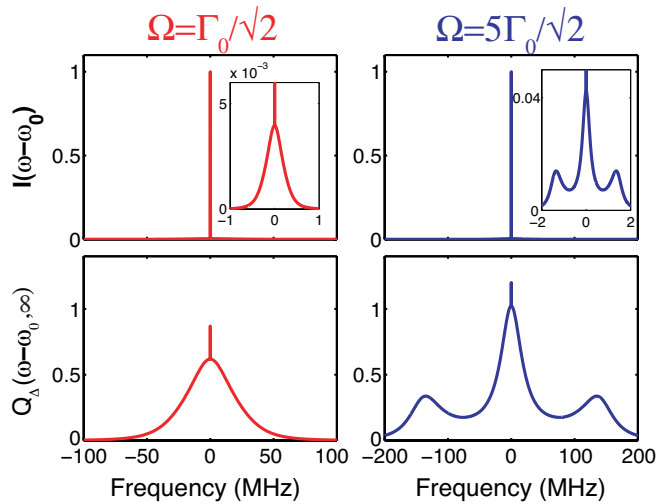}
\caption{Predicted normalized line shape 
(top panes) and 
$Q_{\Delta}(\omega-\omega_0,t=\infty)$ (bottom panes) for a 2-level dye with $\Gamma_0/2\pi=40MHz$ and two different Rabi
frequencies $\Omega$ as indicated.  $\Delta = 0.2 \;\mathrm{MHz}$.  Insets truncate the y-axis to fully display the lineshape outside the vicinity of the high coherent peak at $\omega=\omega_0$.}
\label{fig:1}
\end{figure}
\begin{figure}[t]
\centering
\includegraphics[width=3.0in]{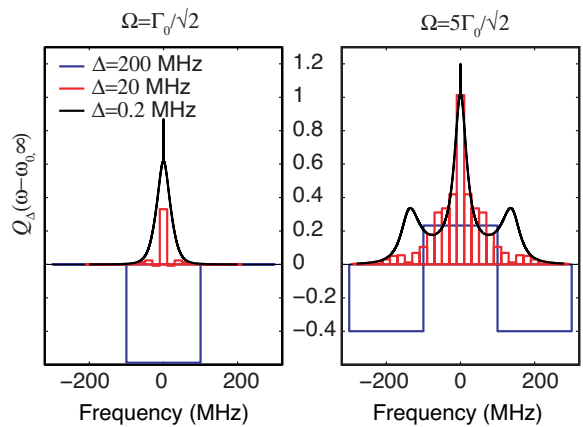}
\caption{$ Q_{\Delta}(\omega,t=\infty)$ for different values of $\Delta$ and the
same cases as in fig. \protect{\ref{fig:1}}.  The bar graph format is used to emphasize the 
size of and location of the frequency bins, but is absent at the finest discretization for
clarity. }
\label{fig:2}
\end{figure}

A traditional measure of broadband photon statistics is Mandel's Q
parameter \cite{optlett}, which is defined as the ratio of the second factorial
cumulant of $N_{(\omega,\infty)}(t) \equiv N(t)$ to the first factorial cumulant (i.e the average)
of $N(t)$.  
$Q \left(t\right)\equiv  \left [ \langle N^2 (t) \rangle - \langle N (t) \rangle^2 - \langle N (t) \rangle \right ] /  \langle N (t) \rangle$.
We introduce a generalization of this quantity appropriate to photon counting within a finite size frequency window
\begin{eqnarray}
Q_{\Delta}(\omega,t) &=&  \frac{ \langle N^{(2)}_{(\omega,\Delta)} (t) \rangle - \langle N_{(\omega,\Delta)} (t) \rangle^2 }{ \langle N_{(\omega,\Delta)} (t) \rangle} \\
&=&  \frac{ \langle N^2_{(\omega,\Delta)} (t) \rangle - \langle N_{(\omega,\Delta)} (t) \rangle^2 - \langle N_{(\omega,\Delta)} (t) \rangle}{ \langle N_{(\omega,\Delta)} (t) \rangle}
\nonumber
\end{eqnarray}
and $Q_{\infty}\left(\omega,t\right)=Q\left(t\right)$.  Fig. \ref{fig:1} plots both the emission lineshape (with finite resolution) $I(\omega-\omega_0) \equiv \lim_{t\rightarrow \infty} \frac{d}{dt} \langle N_{(\omega,\Delta)} (t) \rangle$ and $Q_{\Delta}(\omega,\infty)$ for resonant excitation conditions $(\omega_L = \omega_0)$ and $\Delta = \Gamma_0 /400\pi=0.2\;\mathrm{MHz}$.  Two
different values of the Rabi frequency are considered: $\Omega=\Gamma_0/\sqrt{2}$ and $\Omega=5\Gamma_0/\sqrt{2}$.  The effect of frequency binning is barely discernible in the lineshape when $\Delta$ is chosen so small.  
Our results are essentially identical to the classical emission spectrum of Mollow \cite{mollow}, excepting the delta function ``coherent" \cite{mollow,ct} contribution at $\omega=\omega_L$, which adopts a finite height after frequency binning.  Plots for $Q_{\Delta}\left(\omega,\infty\right)$ have not been reported previously, and at first sight our results appear surprising.  The values selected for $\Omega$ in the
chosen examples both yield sizable negative values for the traditional broadband $Q$ parameter ($-3/4$ and $-0.11$ for $\Omega=\Gamma_0/\sqrt{2}$ and $\Omega=5\Gamma_0/\sqrt{2}$ respectively), however $Q_{\Delta}(\omega,\infty)$ is seen to be positive over the entire frequency axis.  The implication is that the antibunching phenomenon associated with $Q<0$ is
due to correlations between photons of different frequencies.  To make this point more explicitly, we plot $Q_{\Delta}(\omega,\infty)$ for different choices of $\Delta$ in fig. \ref{fig:2}. 
$Q_{\Delta}(\omega,\infty)$ is seen to become negative  over portions of the frequency axis as $\Delta$ 
approaches the width of the peaks in the spectrum. Related behavior has been predicted for
the intensity correlation function  ($g2$)  of photons originating from a single well-resolved
sideband in the Mollow triplet \cite{CT2,Nienhuis,ct}. In that case, antibunching may be 
explained via the allowed sequence of photon emissions in the radiative cascade predicted
by the dressed atom picture \cite{ct}.  Interestingly, narrowband bunching has previously been
attributed to properties of the detector \cite{Nienhuis}, but we find the same effect  in our
observables that focus solely on photon emission.

\begin{figure}[t]
\centering
\includegraphics[width=3.4in]{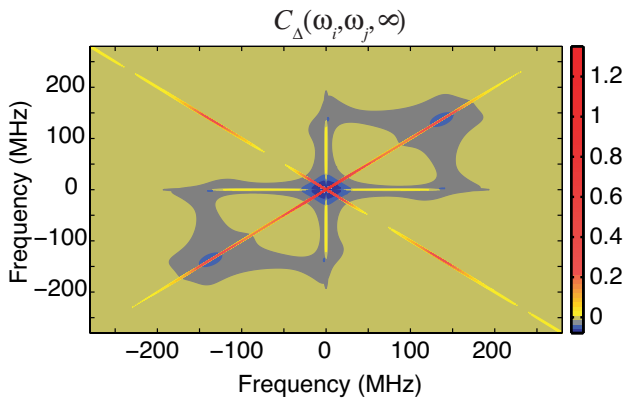}
\caption{Contour plot of the normalized factorial covariance function 
(see eq. \protect{\ref{eq:covar}}).  We consider the $\Omega=5\Gamma/\sqrt{2}$ 
case of figs. \protect{\ref{fig:1}} and \protect{\ref{fig:2}} and have set $\Delta = 2\; \mathrm{MHz}$.}
\label{fig:3}
\end{figure}
Eq. \ref{eq:moms} is easily generalized to calculate correlations between photons at different frequencies.  We define a normalized photon covariance function as:
\small
\begin{align}
\label{eq:covar}
& C_{\Delta}(\omega_i,\omega_j,t) = \\
&  \frac{ \langle N_{(\omega_i,\Delta)}(t)  N_{(\omega_j,\Delta)}(t)\rangle - \langle N_{(\omega_i,\Delta)}(t)\rangle \langle N_{(\omega_j,\Delta)}(t)\rangle }
{\sqrt{ \langle N_{(\omega_i,\Delta)}(t) \rangle \langle N_{(\omega_j,\Delta)}(t) \rangle}} - \delta_{\omega_i,\omega_j}\nonumber.
\end{align} 
\normalsize
A discretized version of this correlation function in the limit $t \rightarrow \infty$ is plotted in fig. \ref{fig:3}, where $\omega_{i(j)}$ 
have been chosen to follow $\omega_i = \omega_0 + r \Delta$, where  $r$ is any integer.  When
$\omega_i = \omega_j$, $C_{\Delta}(\omega_i,\omega_i,t)=Q_{\Delta}(\omega_i,t)$.  Otherwise, $C_{\Delta}(\omega_i,\omega_j,t)$ simply
represents the covariance in photon number, normalized so as to give a finite result in the long time limit.
Fig. \ref{fig:3} demonstrates that although $Q_{\Delta=\Gamma_0/40\pi}(\omega_i,\infty)\geq0$, the total $Q$ parameter is dominated by negative contributions from photons that are well separated in frequency; broadband measurement
of $Q$ contains important contributions from correlations spanning the entire spectrally active region of
the transition.
The positive inter-sideband peaks in fig. \ref{fig:3} reflect the correlated emission of photons from opposite sidebands. This is in qualitative agreement with the inter-sideband bunching expected for a 2-level system excited far from resonance \cite{CT2}. The phenomenon is attributable to
the necessary paring of photons from the two sidebands in order to maintain total energy conservation
as photons of energy $\hbar \omega_L$ are absorbed by the molecule.

Our treatment of photon emission statistics is general and relies on no approximations beyond
the RWA and Markov assumption for system dynamics.  It is valid for arbitrary field strengths
and does not assume particular physical regimes for the molecular system. Moreover, the present approach
provides photon correlations between all possible frequency pairs, which enables calculation
for any possible detector bandwidth and a quantitative demonstration of how seemingly
inconsistent broadband versus narrowband statistics can arise from the same physical phenomena.
This framework
should prove valuable in the interpretation of future SMS experiments where moments higher than 1 will be measured and in understanding
the molecular dynamics that such measurements probe.  Several 
multi-state dye models are discussed in ref. \cite{bel} and will be treated in a future study \cite{inprep}.

This work was supported by the NSF (CHE-0349196).  F. B. is an
Alfred P. Sloan Research Fellow and a Camille Dreyfus Techer-Scholar. 
%



\begin{thebibliography}{19}
\bibitem[{not()}]{note}
\bibinfo{note}{We assume the electronic splitting far exceeds any nuclear
  energy scales in the system, and have replaced $\omega_k$ within the standard
  expression for $\gamma$ with $\omega_0$. Our assumptions (dipole operator
  completely non-diagonal in the electronic space and large electronic
  splitting) insure that $\omega_k \sim \omega_0$ for all relevant photons.}
\end{thebibliography}
\end{document}